\documentclass[aps,orb,twocolumn,amssymb,showpacs]{revtex4}

\usepackage[dvipdf]{epsfig}

\begin{document}
\def\be{\begin{equation}}
\def\ee#1{\label{#1}\end{equation}}

\title{Fermions as sources of accelerated regimes in cosmology}
\author{M. O. Ribas}\email{marlos@fisica.ufpr.br}
\affiliation{Faculdades Integradas Esp\'\i rita, Rua Tobias de
Macedo Jr.  333, 82010-340 Curitiba, Brazil}
\affiliation{Departamento de F\'\i sica, Universidade Federal do
Paran\'a, Caixa Postal 19044, 81531-990 Curitiba, Brazil}
\author{F. P. Devecchi}\email{devecchi@fisica.ufpr.br}
\author{G. M. Kremer}\email{kremer@fisica.ufpr.br}
\affiliation{Departamento de F\'\i sica, Universidade Federal do
Paran\'a, Caixa Postal 19044, 81531-990 Curitiba, Brazil}

\begin{abstract}
In this work it is investigated if  fermionic sources could be
responsible for accelerated periods during the evolution of a
universe where a matter field would answer for the decelerated
period. The self-interaction potential of the fermionic field is
considered as a function of the scalar and pseudo-scalar
invariants. Irreversible processes of energy transfer between the
matter and gravitational fields are also considered. It is shown
that the fermionic field could behave like an inflaton field in
the early universe and as dark energy for an old universe.

\end{abstract}
 \maketitle

%%%%%%%%%%%%%%%%%%%%%%%%%%%%%%%%%%%%%%%%%%%%%%%%%%%%%%%%%%%%%%%%%%%%%%%%%%
\section{Introduction}
%%%%%%%%%%%%%%%%%%%%%%%%%%%%%%%%%%%%%%%%%%%%%%%%%%%%%%%%%%%%%%%%%%%%%%%%%%

The search for constituents responsible for  accelerated periods in
the evolution of the universe is a fundamental topic in cosmology.
Usually  the formulations include elements of general relativity,
field theory and thermodynamics, putting under analysis the
evolution of space-time variables like the scale factor, its
acceleration and energy densities. Several candidates has been
tested for describing both the inflationary period and the present
accelerated era: scalar fields, exotic equations of state and
cosmological constants.

Another possibility is to consider fermionic fields as gravitational
sources for an expanding universe. Fermionic sources has been
investigated using several
 approaches, with results including exact solutions,
anisotropy-to-isotropy scenarios and cyclic cosmologies (see, for
example~\cite{Saha, Armendariz, Obukhov}).

In the present work the connection between general relativity and
the Dirac equation is done via the tetrad formalism, where the
components of the tetrad or ``vierbein" play the role of  the
gravitational degrees of freedom.  The interactions between the
constituents are modeled through the presence of a non-equilibrium
pressure term in the source's energy-momentum tensor. Besides, it is
considered a self-interaction term for the fermionic constituent, in
the form of a potential that can assume several forms (included the
Nambu-Jona-Lasinio case~\cite{NJL}).

While testing  fermionic sources as responsible of  accelerated
periods different regimes are possible. In a young universe scenario
the fermion produces a fast expansion where matter (included via a
barotropic equation of state) is created till it starts to
predominate and the initial accelerated expansion gives place to a
decelerated era, dominated by a matter field, which ends when the
fermionic field predominates again leading to an accelerated era. In
this case the fermionic field plays the role of the inflaton in the
early period of the universe and of dark energy for the old
universe, without the need of a cosmological constant term or a
scalar field. In an old universe scenario an initially matter
dominated period gradually turns into a dark (fermionic) energy
period when an accelerated regime starts and remains for the rest of
the evolution of the system.

 The manuscript is
structured as follows. In section II we make a brief review of the
elements of the tetrad formalism used  to include fermionic and
matter fields in a dynamical curved space-time.  The field equations
for an isotropic, homogeneous and spatially flat universe are
derived in section III . In section IV it is presented the analysis
of the different scenarios in which the fermionic constituent
answers for accelerated eras and the transitions from accelerated to
decelerated periods and vice-versa. Finally, in section V we display
our conclusions. The metric signature used is  $(+,-,-,-)$ and units
have been chosen so that $8\pi G=c=\hbar=k=1$.

%%%%%%%%%%%%%%%%%%%%%%%%%%%%%%%%%%%%%%%%%%%%%%%%%%%%%%%%%%%%%%%%%%%%%%%%%%
\section{Dirac and Einstein equations}
%%%%%%%%%%%%%%%%%%%%%%%%%%%%%%%%%%%%%%%%%%%%%%%%%%%%%%%%%%%%%%%%%%%%%%%%%%

In this section we present briefly the techniques that are used to
include fermionic sources in the Einstein theory of gravitation and
for a more detailed analysis the reader is referred to
\cite{Wald,Ryder,Weinberg,Bir}. The starting point is that the gauge
group of general relativity does not admit a spinor representation
and the tretad formalism is invoked to solve the problem. Following
the general covariance principle, a connection between the tetrad
and the metric tensor $g_{\mu\nu}$ is established through the
relation
\begin{equation}
 g_{\mu\gamma} =e^a_\mu e^b_\gamma\eta_{ab},\qquad a
=0,1,2,3
 \label {1}
\end{equation}
 where $e^a_\mu$ denotes the tetrad or  ``vierbein" and $\eta_{ab}$ is the Minkowski metric tensor.
 Here Latin indices refer to the
 local  inertial frame  whereas Greek indices to the  general system.

 As it was said above, the main objective of this work is  to describe the behavior of fermions
 in the presence of a gravitational field, and the next step is to construct an action for this system.
The Dirac lagrangian density in  Minkowski space-time is
\begin{equation}
L_D=\frac{\imath}{2}[
\overline\psi\gamma^a\partial_a\psi-(\partial_a\overline\psi)\gamma^a\psi]-m\overline\psi\psi-V,
\label{2}
\end{equation}
where the spinors are treated as classically commuting
fields~\cite{Ryder}. Above, $m$ is the fermionic mass,
$\overline\psi$$=$$\psi^\dag$$\gamma^0$ denotes the adjoint spinor
field and the term $V$, which is a function of $\psi$ and
$\overline\psi$,  describes the potential density of
self-interaction between fermions. The general covariance principle
imposes that the Dirac-Pauli matrices $\gamma^a$ must be replaced by
their generalized counterparts $\Gamma ^{\mu}=e^\mu_a\gamma^a$,
whereas the generalized Dirac-Pauli matrices satisfy the  extended
Clifford algebra, i.e., $ \{\Gamma^\mu,\Gamma^\nu\}=2g^{\mu\nu}.$

 In a second step we need to substitute the ordinary derivatives by their covariant versions
\begin{equation}
\partial_\mu\psi\rightarrow D_\mu\psi= \partial_\mu\psi-\Omega_\mu\psi,\quad
\partial_\mu\overline\psi\rightarrow
D_\mu\overline\psi=\partial_\mu\overline\psi+\overline\psi\Omega_\mu,
 \label{3}
\end{equation}
where  the spin connection $\Omega_\mu$ is given by
\begin{equation}
\Omega_\mu=-\frac{1}{4}g_{\mu\nu}[\Gamma^\nu_{\sigma\lambda}
-e_b^\nu(\partial_\sigma e_\lambda^b)]\gamma^\sigma\gamma^\lambda,
\label{4}
\end{equation}
with $\Gamma^\nu_{\sigma\lambda}$ denoting the Christoffel symbols.
Hence, the generally covariant Dirac lagrangian becomes
\begin{equation}
L_D=\frac{\imath}{2}[ \overline\psi\,\Gamma^\mu
D_\mu\psi-(D_\mu\overline\psi)\Gamma^\mu\psi]-m\overline\psi\psi-V.
\label{5}
\end{equation}

  The field equations are obtained from the  total action
\begin{equation} S(g,\psi,\overline\psi)=\int \sqrt{-g}\,L_t\,d^4 x,\label{6}
\end{equation}
 where  $L_t=L_g+L_D+L_m$ is the total lagrangian density.
 $L_g=R/2 $, with $R$ denoting the curvature scalar
 is the gravitational Einstein lagrangian density for fermions
 which are  minimally coupled to the gravitational field. $L_D$   is the Dirac lagrangian  density
 (\ref{5}) and $L_m$  is the lagrangian density of the matter field.

  From the lagrangian density (\ref{5}), through Euler-Lagrange equations, we
  obtain the Dirac equations for the spinor field  and its adjoint coupled to the gravitational field
\begin{equation}
\imath\Gamma^\mu D_\mu\psi-m\psi-{dV\over d{\overline\psi}}= 0,\quad
\imath D_\mu\overline\psi\,\Gamma^\mu+m\overline\psi+{dV\over
d\psi}= 0. \label{7}
\end{equation}

 The variation of the action (\ref{6}) with respect to the tetrad
 leads to Einstein field equations
\begin{equation}
R_{\mu\nu}-\frac{1}{2}g_{\mu\nu}R=-T_{\mu\nu},\label{8}
\end{equation}
 where  $T_{\mu\nu} $ is the total energy-momentum tensor which
 is a sum of the energy-momentum tensor of the fermionic field
 $T^{\mu\nu}_f$ and of the matter field $T^{\mu\nu}_m$, i.e.,
$T^{\mu\nu}=T^{\mu\nu}_m+T^{\mu\nu}_f$. Furthermore, the symmetric
form of the energy-momentum tensor of the fermionic field is given
by
 \[
 T^{\mu\nu}_f=\frac{\imath}{4}\left[\overline\psi\Gamma^\mu
 D^\nu\psi+\overline\psi\Gamma^\nu D^\mu\psi\right.
\]
\begin{equation}
\left.
 -D^\nu\overline\psi
 \Gamma^\mu\psi
 -D^\mu\overline\psi\Gamma^\nu\psi\right]
 -g^{\mu\nu}L_D.
 \label{9}
 \end{equation}

%%%%%%%%%%%%%%%%%%%%%%%%%%%%%%%%%%%%%%%%%%%%%%%%%%%%%%%%%%%%%%%%%%%%%%%%%%
\section{Field Equations}
%%%%%%%%%%%%%%%%%%%%%%%%%%%%%%%%%%%%%%%%%%%%%%%%%%%%%%%%%%%%%%%%%%%%%%%%%%

 The  Robertson-Walker metric incorporates the
homogeneity and isotropy hypotheses for the universe. Here we
consider a spatially flat universe described by the metric
\begin{equation}
ds^2=dt^2-a(t)^2(dx^2+dy^2+dz^2), \label{11}
\end{equation}
where $a(t)$ refers to the cosmic scale factor.

The total energy-momentum tensor for an isotropic and homogeneous
universe -- which is composed by  fermionic and matter fields and
where dissipative processes are taken into account -- is written as
 \be
 {(T^\mu}_\nu)={\rm diag}(\rho,-p-\varpi,-p-\varpi,-p-\varpi).
 \ee{12}
 Above, the total energy density $\rho$ and the total pressure $p$ are given
 as a sum of the corresponding terms of the fermionic and matter fields, i.e., $\rho=\rho_f+\rho_m$ and
 $p=p_f+p_m$. Moreover, the quantity $\varpi$ refers to a non-equilibrium
 pressure which is related to dissipative processes during the
 evolution of the universe and represents an irreversible process of energy transfer between
 the matter and the gravitational field~\cite{KD}.

 Thanks to the Bianchi identities, the covariant differentiation of Einstein field equations (\ref{8}) leads to
 the conservation law of the total energy-momentum tensor
 ${T^{\mu\nu}}_{;\nu}=0$, hence  it follows by using the representation (\ref{12})
 the evolution equation for the total energy density:
 \be
 \dot\rho+3H(\rho+p+\varpi)=0,
 \ee{13}
 where the dot refers to a differentiation with respect to time and $H=\dot a(t)/a(t)$ denotes the Hubble parameter.
Furthermore, from Einstein field equations (\ref{8}) it follows the
 Friedman and acceleration equations
 \be
 H^2={1\over3}\rho,\qquad {\ddot a\over a}=-{1\over 6}(\rho+3p+3\varpi),
 \ee{14}
respectively. Only two  equations from (\ref{13}) and (\ref{14}) are
linearly independent.

 For the metric (\ref{11}) the tetrad components read
\begin{equation}
e_0^\mu=\delta_0^\mu,\qquad  e_i^\mu=\frac{1}{a(t)}\delta_i^\mu.
\label{15} \end{equation}
 Also,  the Dirac matrices become
\begin{equation}
\Gamma^0=\gamma^0,\quad  \Gamma^i=\frac{1}{a(t)}\gamma^i, \quad
\Gamma^5=-\imath\sqrt{-g}\,\Gamma^0\Gamma^1\Gamma^2\Gamma^3=\gamma^5
\label{16},
\end{equation}
  from which the spin connection components are obtained, yielding
\begin{equation}
\Omega_0=0,\qquad  \Omega_i=\frac{1}{2}\dot
a(t)\gamma^i\gamma^0.\label{17}\end{equation}

 For an isotropic and homogeneous universe the fermionic field  is an exclusive function of time. So
the Dirac equations (\ref{7}) become
 \be
 \dot\psi+{3\over2}H\psi+\imath m\gamma^0\psi+\imath\gamma^0{dV\over
 d{\overline\psi}}=0,
 \ee{18a}
 \be
 \dot{\overline{\psi}}+{3\over2}H\overline\psi-\imath m\overline{\psi}\gamma^0-\imath{dV\over
 d{\psi}}\gamma^0=0,
 \ee{18}
 thanks to equations (\ref{16}) and (\ref{17}).

The non-vanishing components of the energy-momentum tensor of the
fermionic field follow from (\ref{9}) together with (\ref{5}),
(\ref{16}) -- (\ref{18}), yielding
 \be
 {{(T_f)}^0}_0=m(\overline\psi\psi)+V,
 \ee{19a}
 \be
  {{(T_f)}^1}_1={{(T_f)}^2}_2={{(T_f)}^3}_3=V-{dV\over
 d{\psi}}{{\psi}\over2}-{{\overline\psi}\over2}{dV\over
 d{\overline\psi}},
 \ee{19}
 which are only functions of $\psi$ and $\overline\psi$.
 By identifying the components of the energy-momentum tensor of
 the fermionic field as
 ${{(T_f)}^\mu}_\nu={\rm diag}(\rho_f,-p_f,-p_f,$ $-p_f)$,  one can obtain
  from (\ref{18a})  and  (\ref{18}) by the use of (\ref{19a})  and (\ref{19})
  the conservation law for the energy
 density of the fermionic field:
 \be
 \dot \rho_f+3H(\rho_f+p_f)=0.
 \ee{20}
 Hence, the evolution equation for the energy density of the
 fermionic field decouples from the energy density of the matter
 field, and we have from (\ref{13}) and (\ref{20}):
 \be
 \dot \rho_m+3H(\rho_m+p_m)=-3H\varpi,
 \ee{21}
 where the term $-3H\varpi$ could be interpreted as the energy
 density production rate of the matter field (see e.g.~\cite{KD}).

 According to the
 extended (causal or second-order) thermodynamic theory  the non-equilibrium pressure
 $\varpi(t)$ obeys an evolution equation, whose linearized form
 reads
 \be
 \tau\dot\varpi+\varpi=-3\eta H,
 \ee{22}
 where  $\tau$ denotes  a
 characteristic time and $\eta$ is the so-called bulk viscosity coefficient.

%%%%%%%%%%%%%%%%%%%%%%%%%%%%%%%%%%%%%%%%%%%%%%%%%%%%%%%%%%%%%%%%%%%%%%%%%%
\section{Cosmological Solutions}
%%%%%%%%%%%%%%%%%%%%%%%%%%%%%%%%%%%%%%%%%%%%%%%%%%%%%%%%%%%%%%%%%%%%%%%%%%

 In order to analyze cosmological solutions for the field equations
 of the previous section, we have to specify the potential density
 of self-interaction between the fermions $V$.  According to the
Pauli-Fierz theorem $V$  is an exclusive function of the scalar
invariant $(\overline\psi\psi)^2$ and on the pseudo-scalar invariant
$(i\overline\psi\gamma^5\psi)^2$, i.e., $V=V((\overline\psi\psi)^2,
(\imath\overline\psi\gamma^5\psi)^2)$. Here we are interested in
analyzing self-interaction potentials of the form
 \be
 V=\lambda\left[\beta_1(\overline\psi\psi)^2+\beta_2
 (\imath\overline\psi\gamma^5\psi)^2\right]^n,
 \ee{23}
 where the coupling constant $\lambda$ is a non-negative quantity and $n$ is a constant exponent. We shall
 analyze three cases, namely, (i) $\beta_1=1$ and $\beta_2=0$
 where $V$ is a function only of the scalar invariant; (ii) $\beta_1=0$ and
 $\beta_2=1$ where $V$ depends only on the pseudo-scalar invariant and
 (iii) $\beta_1=\beta_2=1$ where $V$ is a combination of the scalar
 and pseudo-scalar invariants. The  Nambu-Jona-Lasinio potential~\cite{NJL}
 is represented by the last case with $n=1$.

 The energy density and the pressure of the fermions for the
 potential (\ref{23}) are given by
 \be
 \rho_f=m(\overline\psi\psi)+\lambda\left[\beta_1(\overline\psi\psi)^2+\beta_2
 (\imath\overline\psi\gamma^5\psi)^2\right]^n,
 \ee{24}
 \be
 p_f=(2n-1)\lambda\left[\beta_1(\overline\psi\psi)^2+\beta_2
 (\imath\overline\psi\gamma^5\psi)^2\right]^n,
 \ee{25}
 respectively, thanks to (\ref{19}).

 We infer from (\ref{25}) that the fermions could be classified according to the value of the exponent
 $n$. Indeed, for  $n\geq1/2$ the fermions represent a matter field with positive pressure
 ($n>1/2$) or a pressureless fluid ($n=1/2$), whereas for $n<1/2$ the
 pressure of the fermions is negative and they could represent
 either the inflaton or the dark energy.

 For massless fermions, the
 pressure is connected with the energy density by a simple  barotropic
 equation of state $p_f=(2n-1)\rho_f$ and it follows from the
 conservation equation (\ref{20}) for the fermions that $\rho_f\propto
 1/a^{6n}$. We shall not analyze this case here, since the behavior of
 the fermionic field does not differ from that of a matter field when $n>1/2$
 or from that of a bosonic field when $n<1/2$. Moreover, for the
 massive case, we shall deal with only the case where the
 fermionic field behaves as inflaton or dark energy, i.e., the case
 where $n<1/2$.

 For the pressure of the matter field we shall adopt a barotropic
 equation of state, i.e., $p_m=w_m\rho_m$ with $0\leq w_m\leq1$.
 Furthermore, the coefficient of bulk viscosity $\eta$ and the
 characteristic time $\tau$ are assumed to be related with the
 energy density $\rho$ by $\eta=\alpha\rho$ and $\tau=\eta/\rho$,
 where $\alpha$ is a constant~\cite{KD}.

 The system of field equations we shall investigate  in order to find the
 cosmological solutions are:

 \noindent (a) the acceleration equation
 \be
  {\ddot a\over a}=-{1\over 6}(\rho_f+\rho_m+3p_f+3p_m+3\varpi);
 \ee{26}

\noindent (b) the evolution equation for the energy density of the
matter field
  \be
 \dot \rho_m+3H(\rho_m+p_m+\varpi)=0;
  \ee{27}

\noindent (c) the evolution equation for the non-equilibrium
pressure
 \be
 \tau\dot\varpi+\varpi=-3\eta H;
 \ee{28}

 \noindent (d) the Dirac equation (\ref{18a}), which in terms of
 the spinor components  $\psi
=(\psi_1,\psi_2,\psi_3,\psi_4)^T$, becomes
\[
\pmatrix{\dot\psi_1\cr\dot\psi_2\cr\dot\psi_3\cr\dot\psi_4\cr}+\frac{3}{2}H
\pmatrix{\psi_1\cr\psi_2\cr\psi_3\cr\psi_4\cr}+\imath
m\pmatrix{\psi_1\cr\psi_2\cr-\psi_3\cr-\psi_4\cr}
\]
\[
-2\imath(\psi_1^\dag\psi_1+\psi_2^\dag\psi_2-\psi_3^\dag\psi_3-\psi_4^\dag\psi_4)
\pmatrix{\psi_1\cr\psi_2\cr-\psi_3\cr-\psi_4\cr}V'
\]
\be
-2\imath(\psi_3^\dag\psi_1+\psi_4^\dag\psi_2-\psi_1^\dag\psi_3-\psi_2^\dag\psi_4)
\pmatrix{\psi_3\cr\psi_4\cr-\psi_1\cr-\psi_2\cr}V^\star=0. \ee{29}
In equation (\ref{17}) we have introduced the following
abbreviations \be V'=\frac{\partial
V}{\partial(\overline\psi\psi)^2},\qquad V^\star=\frac{\partial
V}{\partial(\overline\psi\gamma^5\psi)^2}. \ee{30}

Equations (\ref{26}) through (\ref{29}) consist of a system of seven
coupled ordinary differential equations for the fields  $a(t)$,
 $\rho_m(t)$,  $\varpi(t)$, $\psi_1(t)$, $\psi_2(t)$,
$\psi_3(t)$ and $\psi_4(t)$ and in the following subsections we
shall find solutions of this system of equations for given initial
conditions.

%%%%%%%%%%%%%%%%%%%%%%%%%%%%%%%%%%%%%%%%%%%%%%%%%%%%%%%%%%%%%%%%%%%%%%%%%%
\subsection{Accelerated-decelerated-accelerated regime}
%%%%%%%%%%%%%%%%%%%%%%%%%%%%%%%%%%%%%%%%%%%%%%%%%%%%%%%%%%%%%%%%%%%%%%%%%%

Let us first analyze the case that corresponds to the evolution of
the early universe, where the fermionic field plays at the beginning
the role of an inflaton and the matter is created by an irreversible
process through the presence of a non-equilibrium pressure $\varpi$.
The initial conditions we have chosen for $t=0$ (by adjusting
clocks) are
 \be
 \cases{
 a(0)=1,\quad \psi_1(0)=0.1\,\imath,\quad \psi_2(0)=1,\quad
 \psi_3(0)=0.3,\cr \psi_4(0)=\imath,\quad \rho_m(0)=0,\quad
 \varpi(0)=0.}
 \ee{31}
The last two initial conditions correspond to a vanishing energy
density of the matter field and a vanishing non-equilibrium pressure
at $t=0$. The conditions chosen here characterize qualitatively an
initial proportion between the constituents in the corresponding era
(i.e., in this case we have a predominating fermionic field over the
matter, indicating the initial inflationary state)

Since equation (\ref{26}) is a second-order differential equation we
need to specify an initial condition for $\dot a(0)$. This condition
follows  from the Friedmann equation (\ref{14})$_1$, i.e.,
 \be
 \dot a(0)=a(0)\sqrt{\rho_f(0)+\rho_m(0)\over3},
 \ee{32}
 with $\rho_f(0)$ determined from equation (\ref{24}). However, we
 have to specify some parameters in order to obtain numerical
 solutions of the coupled system of differential equations (\ref{26})
 through (\ref{29}). These parameters are: (a) $\lambda$, $\beta_1$,
 $\beta_2$ and $n$ which define the self-interacting potential
 (\ref{23}); (b) $m$ which is related to the mass of the fermionic
 field; (c) $w_m$ which defines the matter field through its barotropic equation of state
 $p_m=w_m\rho_m$ and (d) $\alpha$ which is connected with the bulk viscosity
 term $\eta=\alpha\rho$. In order to plot  the figures 1 and 2 we have chosen the
 following values for these parameters:
 \be
 \cases{
 \lambda=0.1,\quad \beta_1=\beta_2=1,\quad n=0.3,\cr m=0.01,\quad
 w_m=1/3,\quad \alpha=1.0, \quad \hbox{and}\quad\alpha= 1.2,}
 \ee{33}
 which correspond to a fermionic field with a negative pressure,
 described by a self-interacting potential that depends on the
 scalar and pseudo-scalar invariants and a matter  field of massless
 particles that could describe a radiation field.

 \begin{figure}\begin{center}
\includegraphics[width=6.5cm]{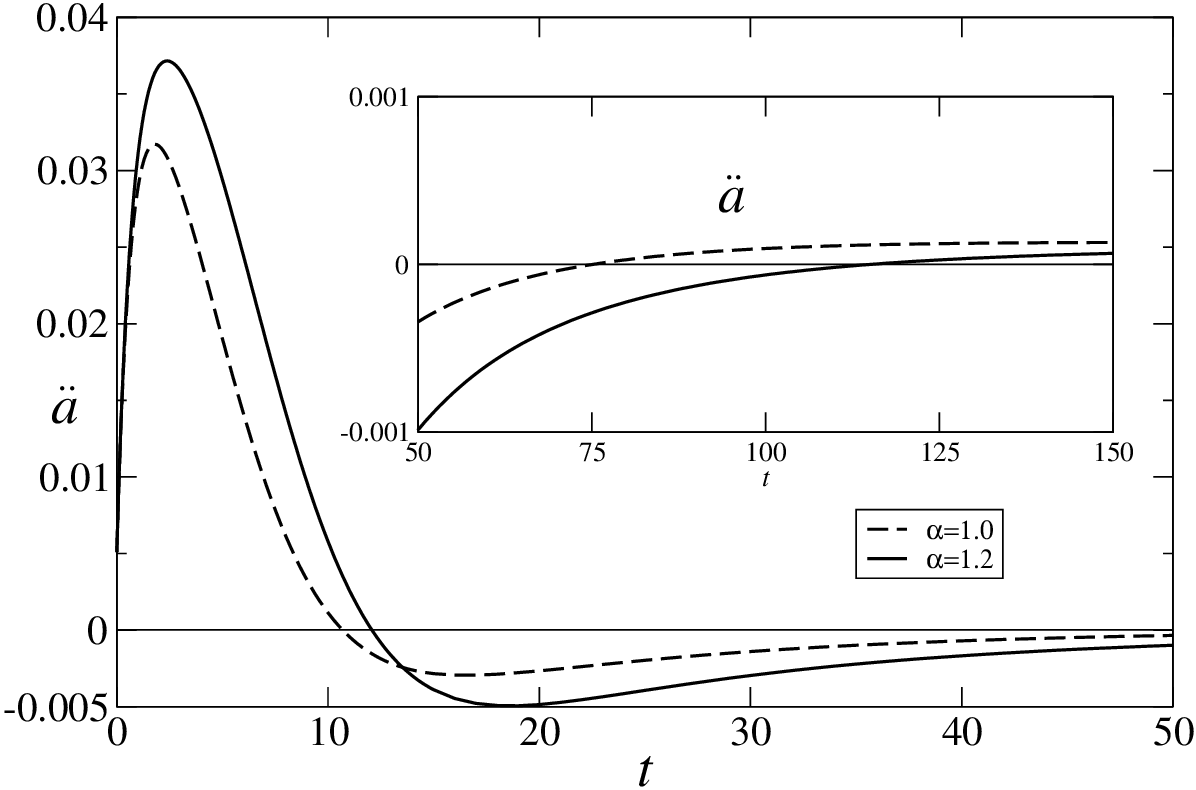}
\caption{Acceleration field $\ddot a$ vs. time $t$.}
\end{center}\end{figure}

 In figure 1 it is plotted the acceleration field $\ddot a$ as function of
 time $t$ for two different values of $\alpha=1.0$ and 1.2, whereas in figure 2 it is
 shown the behavior of the energy densities of the fermionic $\rho_f$ and matter $\rho_m$ fields
 as functions of time $t$. We infer from these figures that there exists an accelerated period
 where the fermionic field dominates and the matter field is created
 at the expenses of an irreversible process of energy transfer between
 the matter and the gravitational field. The accelerated period is
 followed by a decelerated era which is dominated by the matter
 field. Due to fact that the self-interaction potential tends to a constant value for large
 values
 of time, the energy density of the fermionic field overcomes the energy density of the
 matter field and the universe goes into another accelerated period.  It is noteworthy that for the
 bosonic case where the potentials are exponentials or inverse power-laws, one has to add a constant
 value -- which is  similar to introduce a cosmological constant term -- in order to return to an accelerated era
 after the accelerated-decelerated period (see, for
 example~\cite{KDan}). Here the same self-potential interaction
 which  plays the role of an inflaton field at the beginning plays
 the role of  a cosmological constant for later times and could be identify as dark
 energy. We note from the figures that for large values of the
 coefficient $\alpha$ it follows that: (a) the energy density of fermionic field decays more rapidly causing
 a larger accelerated period and (b) the energy density of the matter
 field has a more significant growth and leads to a larger decelerated period.

The same conclusions above could be obtained for a self-interacting
potential which is only a function of the scalar invariant
($\beta_1=1,$ $\beta_2=0$) or of the scalar pseudo-invariant
($\beta_1=0,$ $\beta_2=1$). However, all cases are strongly
dependent on the exponent $n$ of the self-interacting potential and
one can obtain different behaviors  in which there exits only an
accelerated period for the universe or the universe begins with a
decelerated period. This last case will be analyzed in the next
subsection.

It is worth mentioning that the behavior of the fields found in
our analysis is not restricted to the initial conditions given in
(\ref{31}) and the values for the parameters given in (\ref{33}).
 In fact, these represent
 typical values that describe qualitatively the transitions under
 investigation in this case, i.e., the accelerated-decelerated case when
 matter emerges as the predominating constituent and a second
 situation where dark energy, represented by the fermionic field, overcomes
 the matter field and leads to a final accelerated period.

\begin{figure}\begin{center}
\includegraphics[width=6.5cm]{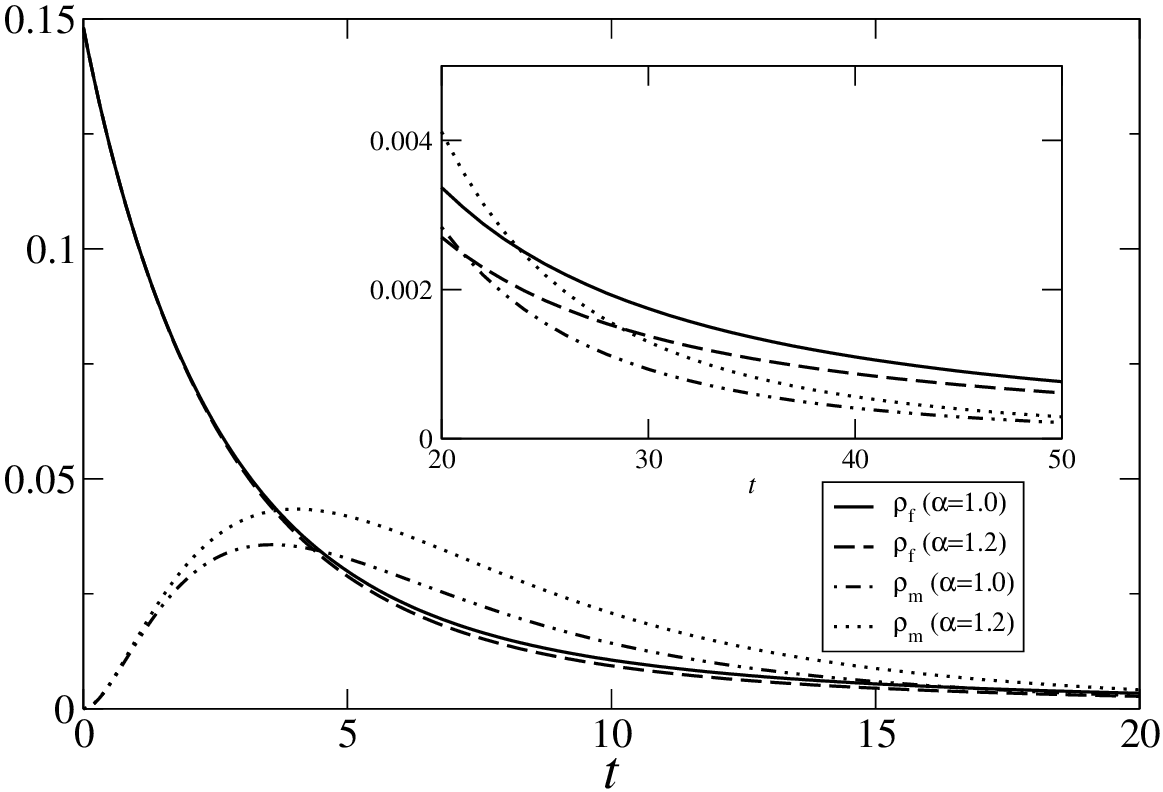}
\caption{Energy densities of  fermionic $\rho_f$ and matter $\rho_m$
fields vs. time $t$.}
\end{center}\end{figure}

%%%%%%%%%%%%%%%%%%%%%%%%%%%%%%%%%%%%%%%%%%%%%%%%%%%%%%%%%%%%%%%%%%%%%%%%%%
\subsection{Decelerated-accelerated regime}
%%%%%%%%%%%%%%%%%%%%%%%%%%%%%%%%%%%%%%%%%%%%%%%%%%%%%%%%%%%%%%%%%%%%%%%%%%

Let us now investigate an old universe dominated by
non-relativistic matter or dust ($w_m=0$ so that $p_m=0$) where a
fermionic field is present but with an energy density  smaller
than that of the matter field. We shall consider the same initial
conditions (by adjusting clocks) as those of previous subsection
for $a(0)$ and $\psi_1(0)$ through $\psi_4(0)$, but for energy
density of the matter field  we regard it as twice that of the
fermionic field, i.e., $\rho_m(0)=2\rho_f(0)$. With respect to the
parameters, we shall choose the same values as above for
$\lambda$, $\beta_1$, $\beta_2$, $m$ and $n$, but different values
for $\alpha$, namely $\alpha=0.1$ and 0.05.  We have also
considered the case where the irreversible processes during the
evolution of the universe are absent. For this last case the
evolution equation of the non-equilibrium pressure (\ref{28}) was
not taken into account. The acceleration and the energy densities
as functions of time are plotted in figures 3 and 4, respectively.
We observe from these figures that there exists a transition from
a high deceleration -- where the universe is dominated by the
non-relativistic matter field -- to a small acceleration --  where
the universe is dominated by the fermionic field which plays the
role of dark energy. By considering irreversible processes there
is no sensitive change in the energy density of the fermionic
field, but the matter field decays more slowly so that the
accelerated period begins later.

The same conclusions of last subsection regarding the
self-interaction potential are also valid for this case.

\begin{figure}\begin{center}%\vskip0.1cm
\includegraphics[width=6.5cm]{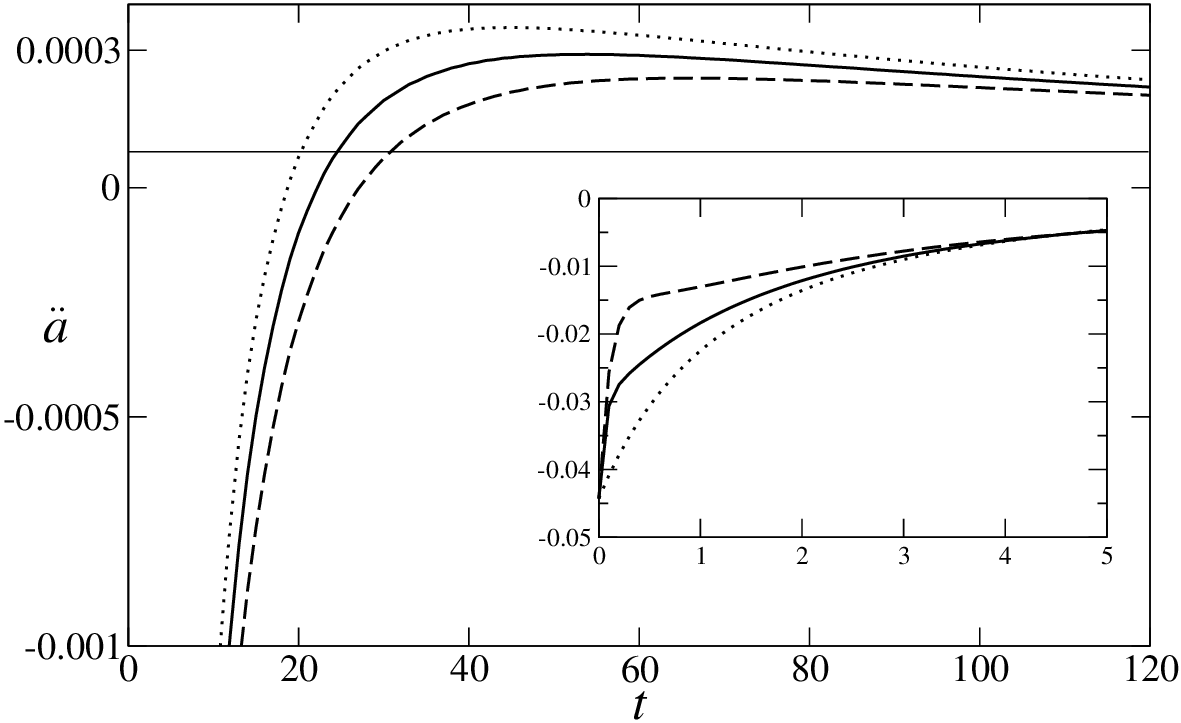}%\vskip0.5cm
\caption{Acceleration field $\ddot a$ vs. time $t$. Dashed line --
$\alpha=0.1$; straight line -- $\alpha=0.05$; dotted line -- without
irreversible processes.}
\end{center}\end{figure}

\begin{figure}\begin{center}
\includegraphics[width=6.5cm]{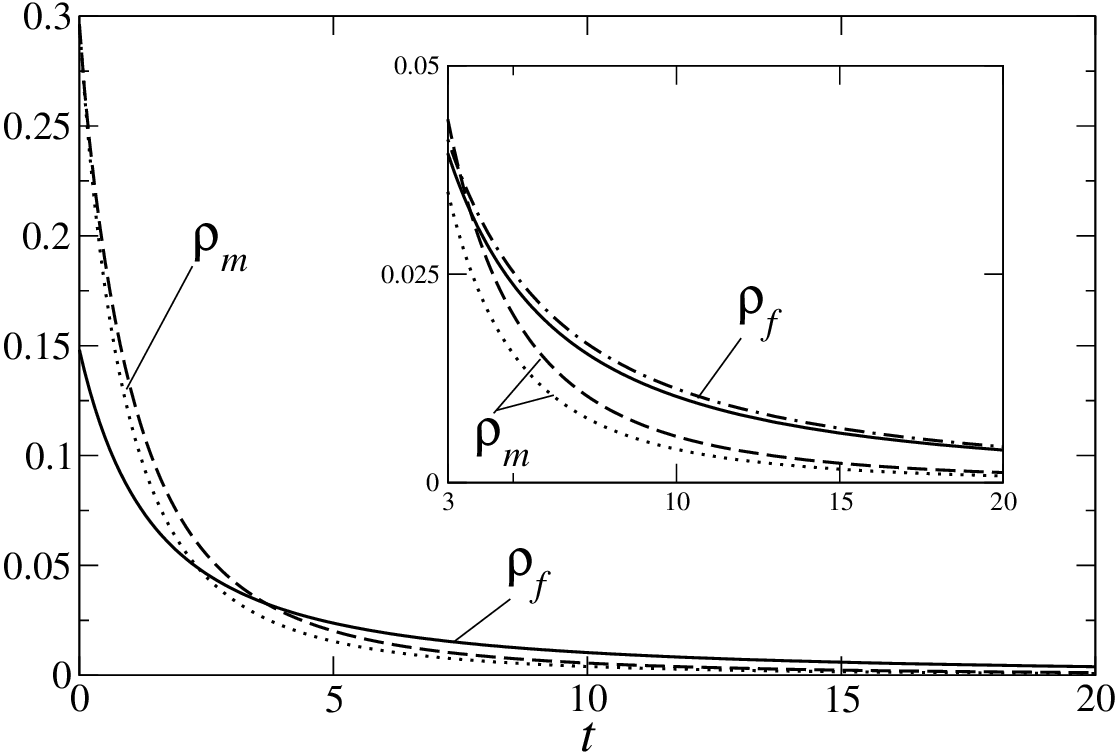}
\caption{Energy densities of  fermionic $\rho_f$ and matter $\rho_m$
fields vs. time $t$. Dotted and dotted-dashed lines -- without
irreversible processes; straight and dashed lines -- $\alpha=0.1$.}
\end{center}\end{figure}

%%%%%%%%%%%%%%%%%%%%%%%%%%%%%%%%%%%%%%%%%%%%%%%%%%%%%%%%%%%%%%%%%%%%%%%%%%
\section{Final Remarks and Conclusions}
%%%%%%%%%%%%%%%%%%%%%%%%%%%%%%%%%%%%%%%%%%%%%%%%%%%%%%%%%%%%%%%%%%%%%%%%%%

 One question that  could be  formulated is about the importance of the irreversible
 processes during the evolution of the universe. To answer this question one should
 discuss the relevance of the bulk viscosity coefficient in the different periods
 of the universe's evolution. In the above calculations it has been used a dimensionless
 coefficient, but if one goes back to its dimensional expression, it is found that it has
 to be multiplied by a factor proportional to the initial value of the Hubble
 parameter. Since we know that the present value of that
 parameter is very small, this coefficient would be playing an
 important role only for early periods of the universe. On the other hand for the
 late period, although its contribution is very small,
 the presence of the viscosity is necessary for describing  the
 thermodynamic dissipative effects in an expanding universe.

 We have investigated the possibility that a fermionic field -- with a self-interacting
 potential that depends on the scalar and pseudo-scalar invariants -- could
 be the responsible for accelerated regimes in the evolution of the
 universe. We have shown that  the fermionic
 field behaves like an inflaton field for the early universe and  later on as a dark
 energy field,
 whereas the matter field was created by an
 irreversible process connected with a non-equilibrium pressure. Moreover, for
 an old decelerated universe dominated by non-relativistic matter
 the fermionic field plays again the role of dark energy and drives the
 universe to an accelerated regime.

%%%%%%%%%%%%%%%%%%%%%%%%%%%%%%%%%%%%%%%%%%%%%%%%%%%%%%%%%%%%%%%%%%%%%%%%%%

\end{document}